\documentclass[a4paper,11pt]{article}
\usepackage{pos}

\title{High-Energy Neutrino and Gamma Ray Production in Clusters of Galaxies}
 \ShortTitle{Neutrinos and $\gamma$-rays from Clusters}

\author*[a,b]{Saqib Hussain}
\author[a,b]{Giulia Pagliaroli}
\author[c]{Elisabete M. de Gouveia Dal Pino}

\affiliation[a]{Gran Sasso Science Institute, Via Michele Iacobucci, 2, 67100 L'Aquila, Italy.}
\affiliation[b]{INFN-LNGS, I-67100 L'Aquila, Italy.}
\affiliation[c]{Institute of Astronomy, Geophysics and Atmospheric Sciences (IAG), University of S\~ao Paulo (USP), R. do Matão, 1226, 05508-090, S\~ao Paulo, Brazil.}

\emailAdd{saqib.hussain@gssi.it}
\emailAdd{giulia.pagliaroli@lngs.infn.it}
\emailAdd{dalpino@iag.usp.br}

\abstract{We compute the contribution from clusters of galaxies to the diffuse neutrino and $\gamma-$ray background. Due to their unique magnetic-field configuration, cosmic rays (CRs) with energy $\leq10^{17}$ eV can be confined within these structures over cosmological time scales, and generate secondary particles, including neutrinos and gamma-rays, through interactions with the background gas and photons. We employ three-dimensional (3D) cosmological magnetohydrodynamical (MHD) simulations of structure formation to model the turbulent intergalactic and intracluster media. We propagate CRs in these environments using multi-dimensional Monte Carlo simulations across different redshifts (from $z \sim 5$ to $z = 0$), considering all relevant photohadronic, photonuclear, and hadronuclear interactions. We also include the cosmological evolution of the CR sources. We find that for CRs injected with a spectral index $1.5 - 2.7$ and cutoff energy $E_\text{max} = 10^{16} - 10^{17}$~eV, clusters contribute to a substantial fraction to the diffuse fluxes observed by the IceCube and Fermi-LAT, and most of the contribution comes from clusters with $M > 10^{14} \, M_{\odot}$ and redshift $z < 0.3$. We also estimated the multimessenger contributions from the 
 local galaxy cluster.}

\ConferenceLogo{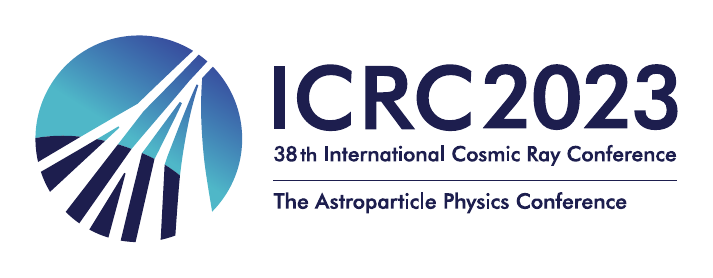}

\FullConference{%
38th International Cosmic Ray Conference (ICRC2023)\\
  26 July - 3 August, 2023\\
  Nagoya, Japan}

\begin{document}
\maketitle

\section{Introdcution}
The diffuse neutrino and $\gamma-$ray background provide a very unique view of the high-energy Universe. The origin of these messengers is unknown. 
%
They may originate from various  astrophysical sources such as galaxy clusters \cite{fang2018linking, murase2013testing},  active galactic nuclei (AGNs) \cite[][see also references therein]{Raniere2022isotropic}, star-forming galaxies (SFGs) \cite[see e.g.,][]{roth2021diffuse}, 
supernova remnants \cite[e.g.,][]{eagle2023fermi}, gamma-ray bursts \cite[see e.g.,][]{lucarelli2022neutrino}.
These high-energy messengers can be produced by a single class of sources i.e., clusters of galaxies \cite{fang2018linking}.

Galaxy Clusters are the largest astrophysical objects of sizes $\sim$~Mpc and magnetic field strength of about $1\,  \mu$~G \cite{nishiwaki2021particle, brunetti2014cosmic}
They can accelerate CRs up to very high-energies $\sim 10^{18}$~eV \cite{brunetti2014cosmic, brunetti2020second} through shock and large-scale turbulence acceleration potentially involving magnetic reconnection.
Due to the large size and magnetic field strength, these structures can confine CRs of energy $\lesssim 10^{17}$~eV up to a few $\sim$~Gyrs \cite{nishiwaki2021particle, ha2023cosmic}. 
%
%
%
While confined, CRs have the ability to interact with gas present within the intercluster medium (ICM) and photon fields, including the cosmic microwave background (CMB) and extragalactic background light (EBL). This interaction leads to the production of high-energy secondary particles such as neutrinos and gamma-rays. Consequently, clusters of galaxies emerge as potential sources for generating high-energy multi-messenger signals \cite{fang2018linking, nishiwaki2021particle}.
%
%
%
Previous analytical and semi-analytical investigations \cite[e.g.,][]{fang2016high,fang2018linking} have estimated the contribution of galaxy clusters to the diffuse neutrino background, revealing their potential to contribute up to $100\%$ of the diffuse neutrino background.
Additionally,  their impact on the diffuse $\gamma-$ray background is also pronounced, particularly for energies exceeding 10 GeV \cite{murase2013testing}.

%

In this work, we summarize our calculations of diffuse backgrounds originating from galaxy clusters, employing the most advanced numerical methods available to date \cite{hussain2021high,hussain2023diffuse}.
In Section \ref{sec:NeuGammClust}, we present our assessment of the contribution of galaxy clusters to the production of neutrinos and $\gamma$-rays, taking into account the entire population of galaxy clusters.  
Following that, in Section \ref{sec:PerseusClust}, we specifically focus on a cluster with properties like the local one, the Perseus cluster. Finally, in Section \ref{sec:discussion}, we discuss our results and draw our conclusions. 

\section{Neutrinos and gamma rays from galaxy clusters}\label{sec:NeuGammClust}

To study the diffuse neutrino and $\gamma-$ray backgrounds we used the most detailed numerical method i.e., combining 3D-MHD simulations with the multi-dimensional Monte-Carlo simulations of test particle propagation using CRPropa code \cite{batista2016crpropa}
%
The background of ICM is probed by 3D-MHD simulations performed by \cite{dolag2005constrained}.
Our assumptions are the following:
(i) CRs consist of only protons;
(ii) neutrinos and $\gamma-$rays are initially produced by purely hadronic interactions inside the clusters; and
(iii) the extragalactic magnetic field is not considered during the propagation of CRs and $\gamma-$rays because it is highly uncertain, and most likely, it will not produce any significant change in the results, especially above $10$~GeV energy.

Our simulation setup has two steps: the first one is to propagate CRs inside clusters 
considering the background magnetic field and density distribution directly from the MHD simulations.
We considered all the relevant CR interactions 
namely: photopion production, Bethe-Heitler pair production, pair production (single, double, triplet), inverse Compton scattering (ICS), and proton-proton (pp) interactions.
We also take into account the adiabatic energy losses due to the expansion of the Universe and the synchrotron losses.
Since the energy of synchrotron photons is most likely $<$~Gev, it is beyond of the scope of this work.
At the end of the first step, we collect the CRs escaped as well as their byproducts $\gamma-$rays and neutrinos, at the edge of the cluster.
In the second step, we propagate those escaped CRs and the $\gamma-$rays from the edge of the cluster to the Earth. 
During CR propagation in the intergalactic medium, we implemented the photopion and Bethe-Heitler pair production due to interactions with CMB and EBL photon fields.
Furthermore, through the propagation of $\gamma-$rays, we accounted for the electromagnetic cascade processes (single, double, triplet pair production, as well as inverse Compton scattering (ICS), which take place both within the clusters and in the intergalactic medium due to CMB and EBL.

\begin{figure*}[htb!]
\centering
\includegraphics[width = 0.7\textwidth]{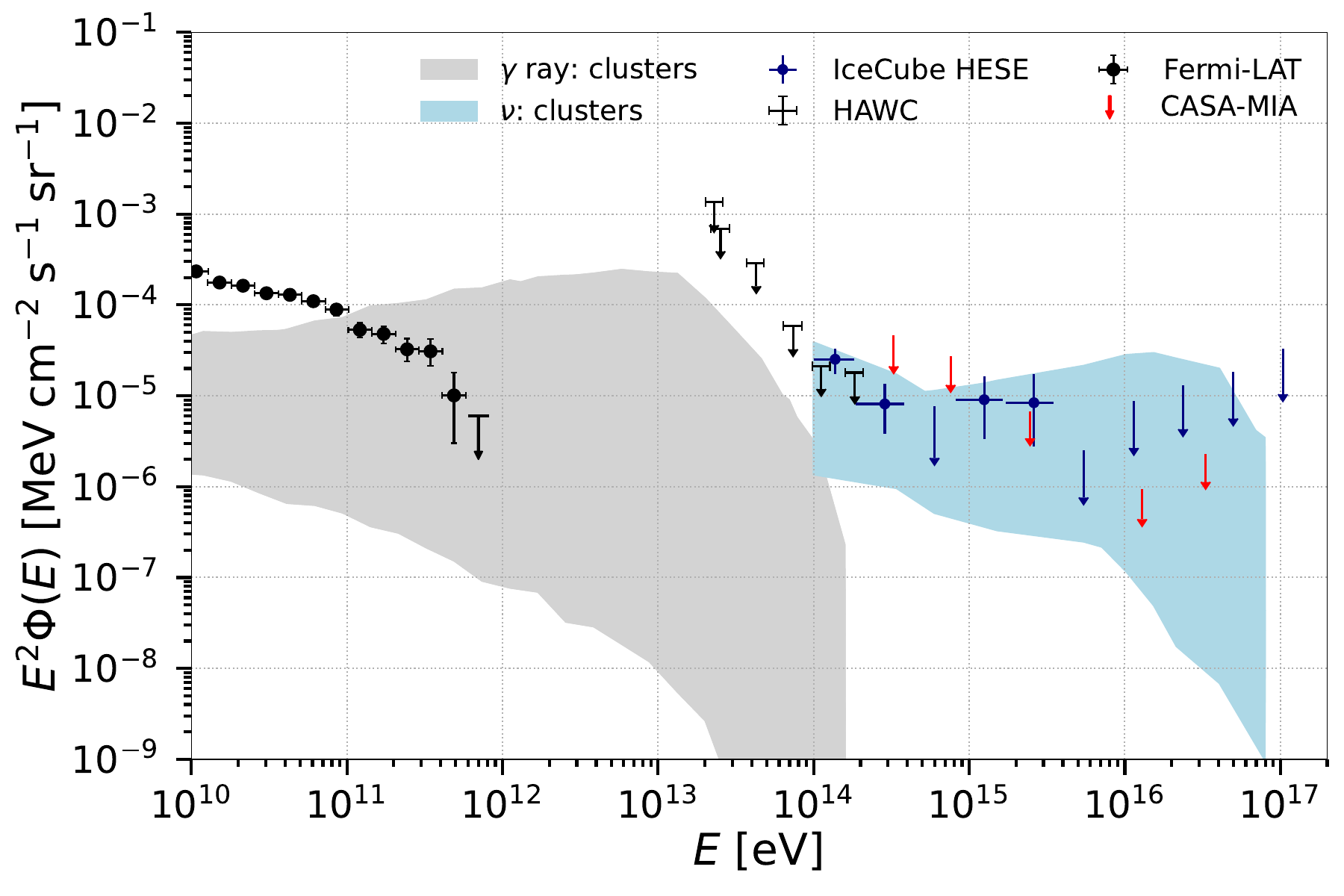}
\includegraphics[width=0.7\textwidth]{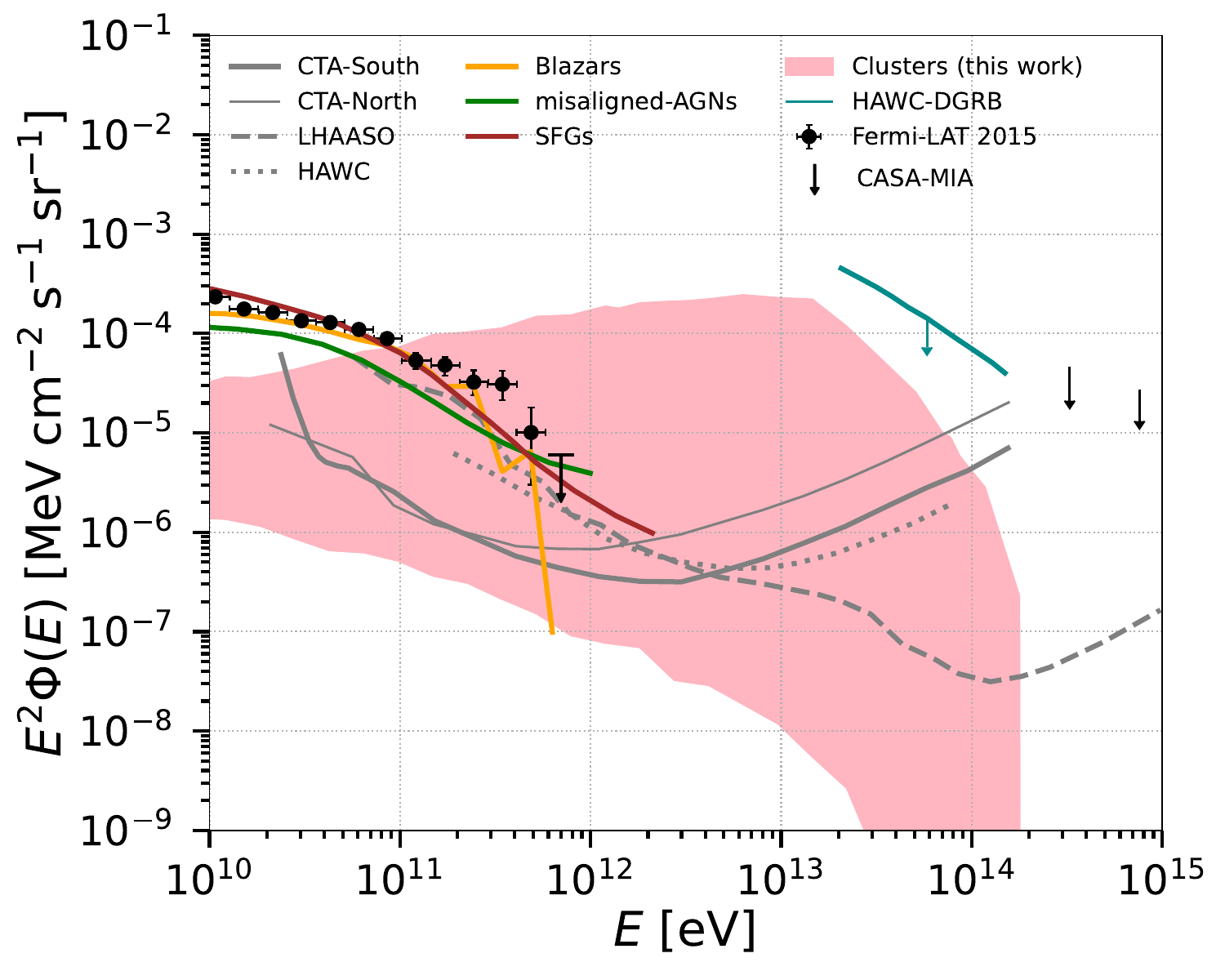}
\caption{Multi-messenger emission from clusters of galaxies.
Neutrino \cite{hussain2021high}  and 
$\gamma-$ray \cite{hussain2023diffuse} from the entire population of galaxy clusters are represented by blue and gray bands, respectively. The neutrino flux is compared with the IceCube data (error bars correspond to the $68\%$ confidence intervals) \cite{aartsen2015evidence}. The $\gamma-$ray flux is compared with the DGRB observed by Fermi-LAT (error bars denote the total uncertainties, statistical and systematic)~\cite{ackermann2015spectrum}, and upper limits from HAWC ($95\%$ confidence level) \cite{HAWC2022limits} and CASA-MIA ($90\%$ confidence level)~\cite{chantell1997limits}. The lower panel compares our $\gamma-$ray flux (pink band) with the
sensitivity curves (gray lines) obtained for point sources from LHAASO \cite{di2016lhaaso}, HAWC \cite{abeysekara2013sensitivity}, and the forthcoming CTA observatories \cite{cta2018science} only for reference purposes.
The contribution from individual sources, namely,  blazars \cite{ajello2015origin}, AGNs \cite{di2013diffuse}, and SFGs \cite{roth2021diffuse} is also shown. 
Extracted from \cite{hussain2021high}  and \cite{hussain2023diffuse}.
}
\label{fig:multi-messenger_cluster}
\end{figure*}

We summarize our results in Fig. \ref{fig:multi-messenger_cluster},  comparing them with the IceCube \cite{aartsen2015searches} and Fermi-LAT data \cite{ackermann2010gev}.
It shows the fluxes of neutrinos and $\gamma-$rays from the entire population of clusters, considering the CR sources embedded in the center.
These fluxes are obtained by injecting CRs in the center of clusters with spectral indices $\alpha=1.5 - 2.5$ and maximum energy in the range $10^{16}-10^{17}$ eV.
The mass range of clusters considered in our simulations is $10^{12}< M/M_{\odot} \lesssim 5 \times 10^{15}$ and the redshift interval is $z\leq 5.0$.
Furthermore, we have assumed that $1\%$  of the cluster luminosity  goes into CRs to be consistent with the prediction of Fermi-LAT~\cite{ackermann2014search}. 
Results obtained from our simulations
are quite comparable with observed diffuse fluxes of neutrinos and $\gamma-$rays.
Fig.~\ref{fig:multi-messenger_cluster} shows the connection between neutrinos and $\gamma-$rays, but it depends on the adopted assumptions such as spectral indices, maximum energy, and the luminosity of CRs \cite{hussain2021high, 
hussain2023diffuse}.

In the lower panel of Fig. \ref{fig:multi-messenger_cluster}, we show the contribution from misaligned AGN \cite{di2013diffuse}, blazars \cite{ajello2015origin}, and SFGs \cite{roth2021diffuse} to diffuse $\gamma-$ray background.
The contribution of individual sources is dominant up to energy $100$ GeV while
the cluster contribution starts to dominate above this energy.
More importantly, our results are comparable with the sensitivity curves of
the High Altitude Water Cherenkov (HAWC) \cite{abeysekara2013sensitivity}, the Large High Altitude Air Shower Observatory (LHAASO) \cite{di2016lhaaso} and the upcoming Cherenkov Telescope Array (CTA) \cite{cta2018science}. 


\section{Multi-messenger from Perseus-like clusters}\label{sec:PerseusClust}

Recently, the Telescope Array (TA) collaboration has observed an excess of CR events of energy $\gtrsim 10^{19.4}$~eV with $\sim 3.5 \; \sigma$ standard deviations toward the center of the Perseus-Pisces supercluster
(PPSC), which is about $75$~Mpc away from the Earth \cite{abbasi2021indications}.
In this section, we focus on the multi-messenger emission including high-energy neutrinos and $\gamma-$rays from an individual cluster with properties similar to  Perseus cluster where there is a high probability of existence of UHECR sources.
We have extracted a single cluster from the global MHD simulation  described in the previous section \cite{hussain2021high,hussain2023diffuse}  to probe the background of a cluster like Perseus with mass $\sim 10^{14.5}\, M_{\odot}$, and studied the propagation of CRs in that medium. We injected CRs at the center of the cluster with spectral index $\alpha=2.3$, $E_\mathrm{max}=10^{17}$~eV, and considered that $1\%$ cluster luminosity goes to CRs.
Initially, neutrinos and $\gamma-$rays are produced by CRs interactions in the ICM.
We have considered all the relevant CR interactions during their propagation both in the cluster and in the intergalactic medium and also
taken into account the $\gamma-$ray cascade inside and outside the cluster, as described above.

\begin{figure}
\centering
\includegraphics[width = 0.6\textwidth]{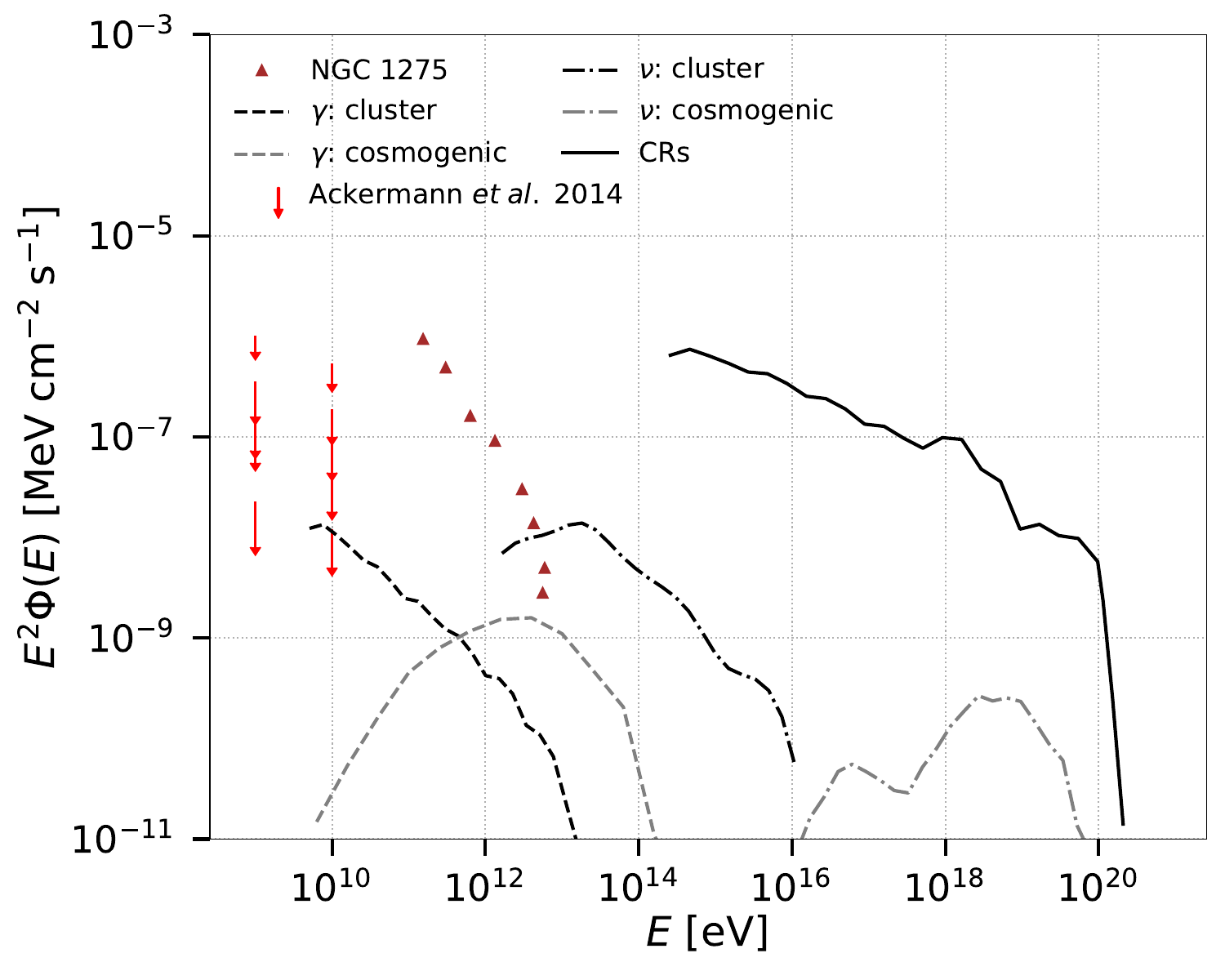}
\caption{Multi-messenger flux of  Perseus-like cluster at $75$ Mpc. Red arrows represent the upper limit from Fermi-LAT \cite{FermiCluster2014search} for five clusters (top to bottom: A400, A3112, A1367, Coma, EXO0422). We also show the high-energy $\gamma-$ray emission from the active galaxy NGC-$1275$ in the center of Perseus cluster, observed by SHALON experiment (1996-2012) \cite{sinitsyna2014emission}. 
}
\label{fig:MMPerseus}
\end{figure}

In Fig.\ref{fig:MMPerseus}, we present the multi-messenger picture of a Perseus-like cluster. 
%
The Fermi-LAT~\cite{ackermann2014search} collaboration estimated the  upper limit for $\gamma-$rays from individual clusters (Fig. \ref{fig:MMPerseus}) and
our results are consistent with their predictions.
Nevertheless, the $\gamma-$ray flux from the central source of Perseus cluster (NGC-$1275$) observed by the SHALON experiment ($1996 - 2012$) \cite{sinitsyna2014emission} is much larger than our results.

\begin{figure}[htb!]
\centering
\includegraphics[width = 0.7\textwidth]{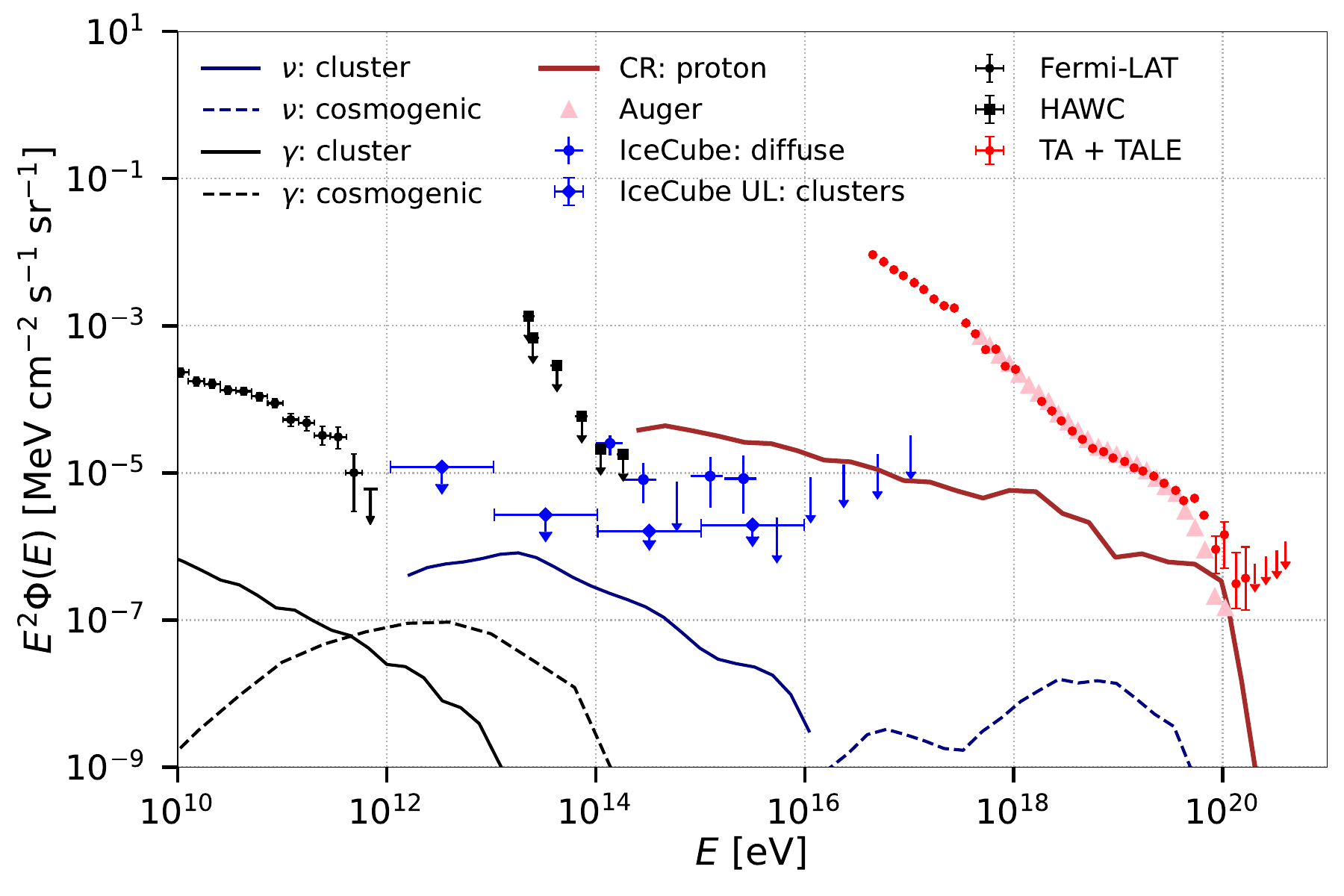}
\caption{Multi-messenger fluxes from an entire population of Perseus-like sources  located at a distance of about $75$ Mpc. The total $\gamma-$ray flux from the entire population is presented by black curves. The DGRB from Fermi-LAT \cite{ackermann2015spectrum}
and the upper limits for DGRB from HAWC-collaboration \cite{HAWC2022limits} are also depicted.
The total neutrino flux from the Perseus-like clusters is presented with blue curves,  the IceCube diffuse flux (blue circle)\cite{aartsen2015searches, aartsen2015evidence}, and the upper limits for clusters (blue diamonds) \cite{abbasi2022searching} is also shown. The red curve shows the spectrum of CRs arrived from Perseus-like sources within a distance $\sim 75$~Mpc and the red marker shows the CR spectrum observed by the telescope array and telescope array low energy extension (TALE ($E\times0.91$))\cite{TACRSpectrum2015}, pink marker represents the Auger data ($E\times1.05$) \cite{batista2019cosmogenic, ICRC342015}.
}
\label{fig:MMObservedPPSC}
\end{figure}

In Fig. \ref{fig:MMObservedPPSC}, we show the total fluxes of CRs, $\gamma-$rays, and neutrinos from  an entire sample of Perseus-like sources in the local Universe.
The number density of clusters with mass $\gtrsim 2\times 10^{14}\, M_{\odot}$ is obtained from our MHD simulation \cite{dolag2005constrained,hussain2021high,hussain2023diffuse}, which is around $N\simeq 10^{-5} \log M \,\text{Mpc}^{- 3}$  and also comparable with \cite{bocquet2016halo}.
Overall, Fig. \ref{fig:MMObservedPPSC} indicates that Perseus-like clusters can significantly contribute to the emissions of UHECRs if their composition consists of protons only. However, 
the CR flux is much smaller than the observed  UHECR flux from Auger data \cite[see e.g., ][also references therein]{halim2022constraining}. 
It is worth noting that the acceleration of CRs depends on rigidity, which suggests that heavier nuclei may not be dominant.
%
%
%
%
%
Nevertheless, it is also important to acknowledge  that individual sources within clusters, such as starburst galaxies and compact objects like magnetars, possess the capability  to accelerate heavier nuclei to extremely high energies. Their inclusion in our analysis could potentially enhance the results for CRs, though the production of neutrinos and $\gamma-$rays is more efficient for protons compared to heavier nuclei. This is primarily because these particles are predominantly produced through pion decay processes.
In the case of heavier nuclei, photodisintegration becomes more prominent than pion production. Therefore, the assumption of considering only protons in our study seems to be reliable \cite[see e.g.,][]{kotera2009propagation}. Still, the contribution of individual sources should be further explored.
In Fig. \ref{fig:MMObservedPPSC} we also showed the flux of secondary particles produced by the interaction of CRs during their propagation in the ICM and intergalactic medium. 
The neutrino flux we obtained for Perseus-like sources is comparable with upper limits recently estimated by the IceCube \cite{abbasi2022searching}.
On the other hand,  the $\gamma-$ray flux is smaller by an order of magnitude from the diffuse flux observed by the Fermi-LAT \cite{ackermann2015spectrum}.

\section{Discussion and Conclusion}\label{sec:discussion}

Our results predict that clusters of galaxies can contribute up to a sizeable percentage  to diffuse neutrinos and $\gamma-$rays.
Our results on neutrinos \cite{hussain2021high} match with the previous studies
\cite{zandanel2015high, fang2016high, fang2018linking, nishiwaki2021particle,abbasi2022searching} and the $\gamma-$rays flux we obtained \cite{hussain2023diffuse} roughly agree with  the predictions of \cite{murase2013testing, zandanel2015high}.

The contribution to diffuse $\gamma-$rays by individual sources such as active galactic nuclei (AGN) \cite{Raniere2022isotropic}, star forming galaxis (SFGs) \cite{roth2021diffuse}, and blazars \cite{ajello2015origin} is dominant over clusters below $100$ GeV energy.
However, above energy $100$ GeV, we showed that the total $\gamma-$ray flux from the entire population of clusters  is dominant over  individual source contribution. 
Therefore, our estimation is extremely important provided that high-energy CRs are present in the clusters.

Our results are  comparable with the diffuse fluxes of neutrinos and $\gamma-$rays observed by IceCube \cite{aartsen2015evidence} and Fermi-LAT \cite{ackermann2015spectrum}, respectively.
Moreover, the $\gamma-$ray flux we obtained from the cluster population up to redshift $z\leq 5.0$ is comparable with the sensitivity curves of the HAWC \cite{harding2019constraints, HAWC2022limits}, the LHAASO \cite{di2016lhaaso}, and even the CTA \cite{cta2018science} which indicates possible observations of $\gamma-$rays from these sources \cite{hussain2021high,hussain2023diffuse}.

Further, we have calculated the multi-messenger emission from Perseus-like sources within a distance of about $75$ Mpc, in our local Universe. 
Our evaluation suggests that the CR flux  from these sources (calculated for protons only) can account for a sizeable percentage of UHECR detected through the Telescope Array (TA).
The current prediction by the TA collaboration of a probable source of UHECRs in the direction of the Perseus cluster provides significance to our findings.
However, our results are not consistent with the Auger data which indicate that the composition of UHECRs mainly consists of heavier nuclei. In the forthcoming work we will account for this contribution. 
%
Nevertheless, the production of neutrinos and $\gamma-$rays calculated from this population of local clusters shows that the flux of neutrino is comparable with the estimated upper limits of IceCube for clusters \cite{abbasi2022searching}, while the $\gamma-$ray flux is about one order of magnitude less than the  Fermi-LAT \cite{ackermann2015spectrum} observations. This indicates that lower mass clusters should also be contributing to this emission, as found in \cite{hussain2023diffuse} (see also Fig. \ref{fig:multi-messenger_cluster}. 

%

We acknowledge that our estimations may be subject to changes if certain parameters are modified, such as the luminosity of the CRs and the spatial distribution of CR sources within clusters. It is important to recognize that the presence of the intergalactic magnetic field can also have minor effects on our results.
However, it is worth noting that the parameters we have chosen for our calculations are in general well-established and similar to those used in previous research. Any modifications to these parameters are expected to have only minor impacts on our overall conclusions.
Moreover, our results establish a clear connection between the three messengers (neutrinos, gamma-rays, and cosmic rays), enabling us to indirectly investigate the properties of CRs within clusters. This connection provides valuable insights into the nature and behavior of CRs in these environments.

\paragraph{Acknowledgments}
This work is partially supported by the  Brazilian agencies FAPESP (grant $2013/10559-5 \, \& \,
17/12828-4)$ and CNPq (grant $308643/2017-8$). The work of SH and GP is partially supported by the research grant number $2017$W$4$HA$7$S ''NAT-NET:Neutrino and Astroparticle Theory Network'' under the program PRIN $2017$ funded by the Italian Ministero dell'Istruzione, dell'Universita' e della Ricerca (MIUR).





\end{document}